\documentclass[conference]{IEEEtran}

\usepackage[T1]{fontenc}
\usepackage[utf8x]{inputenc}
\usepackage{cite}
\usepackage{amsmath,amssymb,amsfonts}
\usepackage{algorithmic}
\usepackage{graphicx}
\usepackage{textcomp}
\usepackage{balance}
\usepackage{xspace}
\usepackage{multirow}
\usepackage{float}
\usepackage{soul}
\usepackage{microtype}
\usepackage{url}
\usepackage{booktabs}
\usepackage{subcaption}

\def\BibTeX{{\rm B\kern-.05em{\sc i\kern-.025em b}\kern-.08em
    T\kern-.1667em\lower.7ex\hbox{E}\kern-.125emX}}

\usepackage[dvipsnames]{xcolor}
\usepackage{listings}

\newcommand\YAMLcolonstyle{\color{red}\mdseries}
\newcommand\YAMLkeystyle{\color{black}\bfseries}
\newcommand\YAMLvaluestyle{\color{blue}\mdseries}

\makeatletter

\newcommand\language@yaml{yaml}

\expandafter\expandafter\expandafter\lstdefinelanguage
\expandafter{\language@yaml}
{
	keywords={true,false,null,y,n},
	keywordstyle=\color{darkgray}\bfseries,
	basicstyle=\YAMLkeystyle,                                 
	sensitive=false,
	comment=[l]{\#},
	morecomment=[s]{/*}{*/},
	commentstyle=\color{purple}\ttfamily,
	stringstyle=\YAMLvaluestyle\ttfamily,
	moredelim=[l][\color{orange}]{\&},
	moredelim=[l][\color{magenta}]{*},
	moredelim=**[il][\YAMLcolonstyle{:}\YAMLvaluestyle]{:},   
	morestring=[b]',
	morestring=[b]",
	literate =    {---}{{\ProcessThreeDashes}}3
	{>}{{\textcolor{red}\textgreater}}1
	{|}{{\textcolor{red}\textbar}}1
	{\ -\ }{{\mdseries\ -\ }}3,
}

\lst@AddToHook{EveryLine}{\ifx\lst@language\language@yaml\YAMLkeystyle\fi}
\makeatother

\newcommand\ProcessThreeDashes{\llap{\color{cyan}\mdseries-{-}-}}

\begin{document}
	
\newcommand{\alessandro}[1]{\textcolor{red}{{\it [Alessandro says: #1]}}}
\newcommand{\oliviero}[1]{\textcolor{green!20!black}{{\it [Oliviero says: #1]}}}
\newcommand{\marco}[1]{\textcolor{orange}{{\it [Marco says: #1]}}}
\newcommand{\leonardo}[1]{\textcolor{green}{{\it [Leo says: #1]}}}

\makeatletter
\newcommand{\linebreakand}{%
\end{@IEEEauthorhalign}
\hfill\mbox{}\par
\mbox{}\hfill\begin{@IEEEauthorhalign}
}
\makeatother

\title{Declarative Dashboard Generation}

\author{\IEEEauthorblockN{Alessandro Tundo}
\IEEEauthorblockA{\textit{University of Milano - Bicocca}\\
Milan, Italy \\
alessandro.tundo@unimib.it}
\and
\IEEEauthorblockN{Chiara Castelnovo}
\IEEEauthorblockA{\textit{University of Milano - Bicocca}\\
Milan, Italy \\
c.castelnovo3@campus.unimib.it}
\and
\IEEEauthorblockN{Marco Mobilio}
\IEEEauthorblockA{\textit{University of Milano - Bicocca}\\
Milan, Italy \\
marco.mobilio@unimib.it}
\linebreakand
\IEEEauthorblockN{Oliviero Riganelli}
\IEEEauthorblockA{\textit{University of Milano - Bicocca}\\
Milan, Italy \\
oliviero.riganelli@unimib.it}
\and
\IEEEauthorblockN{Leonardo Mariani}
\IEEEauthorblockA{\textit{University of Milano - Bicocca}\\
Milan, Italy \\
leonardo.mariani@unimib.it}
}

\maketitle

\begin{abstract}
Systems of systems are highly dynamic software systems that require flexible monitoring solutions to be observed and controlled. Indeed, operators have to frequently adapt the set of collected indicators according to changing circumstances, to visualize the behavior of the monitored systems and timely take actions, if needed. Unfortunately, dashboard systems are still quite cumbersome to configure and adapt to a changing set of indicators that must be visualized. 

This paper reports our initial effort towards the definition of an automatic dashboard generation process that exploits meta-model layouts to create a full dashboard from a set of indicators selected by operators. 
\end{abstract}


\section{Introduction} \label{sec:introduction}

Modern software systems have increasing size and complexity. Consider for instance the Cloud- and Fog-based Systems and Systems of Systems~\cite{lopez2017internet,baresi2019big,colombo2013system} that are operational in many domains, such as Telecommunication, Smart-Cities, Transportation, and Finance. Controlling these systems is extremely challenging, since it requires observing the activity of many services, characterized by a highly dynamic and context-dependent behavior. 

There are already monitoring solutions that allow to collect non-trivial amount of runtime data from many running services, such as the ELK~\cite{elasticsearch2020stack} and Prometheus~\cite{linux2020prometheus} commercial frameworks and the VARYS~\cite{tundo2019varys} and Monasca~\cite{openstack2020monasca} open source research frameworks. While these frameworks can provide end-to-end monitoring capabilities, from data collection to data visualization, they only offer limited autonomous operation capabilities. There are small exceptions, for instance Prometheus may automatically discover new data sources and VARYS can automate probe deployment. However, every time a monitoring system is reconfigured (e.g., a new performance indicator is collected), its dashboard must  be \emph{manually} reconfigured accordingly. In dynamic environments where the set of running services and the set of collected indicators are frequently modified, dashboard reconfiguration can be a major cost and a painful activity to perform. 

Past research focused on the organization of the indicators in a page and across pages. For instance, the notable Goal/Question/Metric approach suggests how to organize measurements across three levels~\cite{caldiera1994goal}. More recent approaches~\cite{janes2013effective,kintz2017creating} provide alternative ways of organizing a set of measurements, for instance considering navigation links across and between groups of indicators. However, there is still little work on the automatic generation of dashboard visualizations starting from a set of indicators collected from a target system. Ideally, dashboards should be (re-)configured using the sole information about the collected metrics, leaving the definition of every aspect concerning presentation to the dashboard configuration engine, alleviating operators from any configuration effort. This can be extremely challenging to achieve since a same set of indicators can be presented in many different ways and taking effective decisions automatically can be hard.

This paper describes an ongoing effort in the direction of defining an \emph{automated dashboard reconfiguration engine} that requires minimal intervention by the operators. In our vision, the dashboard generator should know the semantics of the metrics, to be able to produce relevant visualizations without asking any information beyond the set of collected indicators to operators (e.g., CPU-related indicators might be automatically aggregated in a single visualization by a dashboard). In our solution, the set of concrete visualizations that are generated are controlled by \emph{meta-model layouts}, which are models that define the high-level visualization style that should be adopted. For instance, a meta-model layout may tend to aggregate visualizations in a single page, while a complementary meta-model layout may tend to distribute visualizations over multiple pages that can be navigated. Finally, the resulting set of pages and visualizations must be rendered into a target dashboard tool by generating a suitable set of concrete artifacts (e.g., configuring Kibana~\cite{elasticsearch2020kibana} or Grafana~\cite{grafanalabs2020grafana} requires following different processes and manipulating different artifacts). 

This paper introduces the part of the process that based on a meta-model layout generates a representation of the resulting dashboard that is agnostic with respect to the target dashboard tool. We also describe a core set of meta-model layouts that can be used to define how the resulting dashboard is structured. Rendering the resulting dashboard definition with a target dashboard tool is part of our future work.

\section{Declarative dashboard generation process}
The declarative dashboard generation process is a three-step procedure that creates a ready-to-use dashboard 
from a declarative dashboard definition. 
The first step of the process consists of writing the declarative dashboard definition. The second step consists of transforming the dashboard definition into a technology-agnostic dashboard layout by exploiting a meta-model that defines the arrangement of the individual dashboard items. The third step translates the technology-agnostic dashboard layout into a concrete dashboard for a target dashboard tool.

The next subsections describe each step providing details about the involved objects and the translation mechanisms.

\subsection{Declarative Dashboard Definition}

The declarative dashboard definition step allows clients, which could be both regular \emph{users} and \emph{automated tools}, to define potentially complex dashboards by only providing high-level information. This approach can be used to obtain complex dashboards starting from an input that is as simple as a hierarchy of \textit{Key Performance Indicators (KPIs)} and \textit{Visualizations} objects. We refer to this input as the \emph{declarative dashboard definition}. We envision the possibility to obtain the declarative dashboard definition automatically from an even simpler input, which might consist of just a plain list of KPIs to be monitored, but this is part of our future work.

Listing~\ref{lst:definition} shows a sample declarative dashboard definition. The KPI objects and the Visualizations that are part of the declarative dashboard definition can be \textit{Simple} or \textit{Composed}. 
A \textit{Simple KPI} is defined as a \emph{metric} collected from a \emph{target}, which might be a service or a resource (see for example the simple CPU System KPI defined in Listing~\ref{lst:definition}, lines 2-6). A \textit{Composed KPI} is defined as a \emph{set of KPIs} (either Simple or Composed) (see for example the composed CPU Total KPI in Listing~\ref{lst:definition}, lines 9-11) and a \emph{transformation function} that can be applied to this set of KPIs to obtain a new \emph{derived KPI} (for example the CPU Total is derived from the CPU System and CPU User KPIs using the average function as specified in Listing~\ref{lst:definition}, line 12). 

\begin{figure*}[ht]
\centering
\begin{subfigure}[t]{.32\textwidth}
  \centering
  \includegraphics[width=\linewidth]{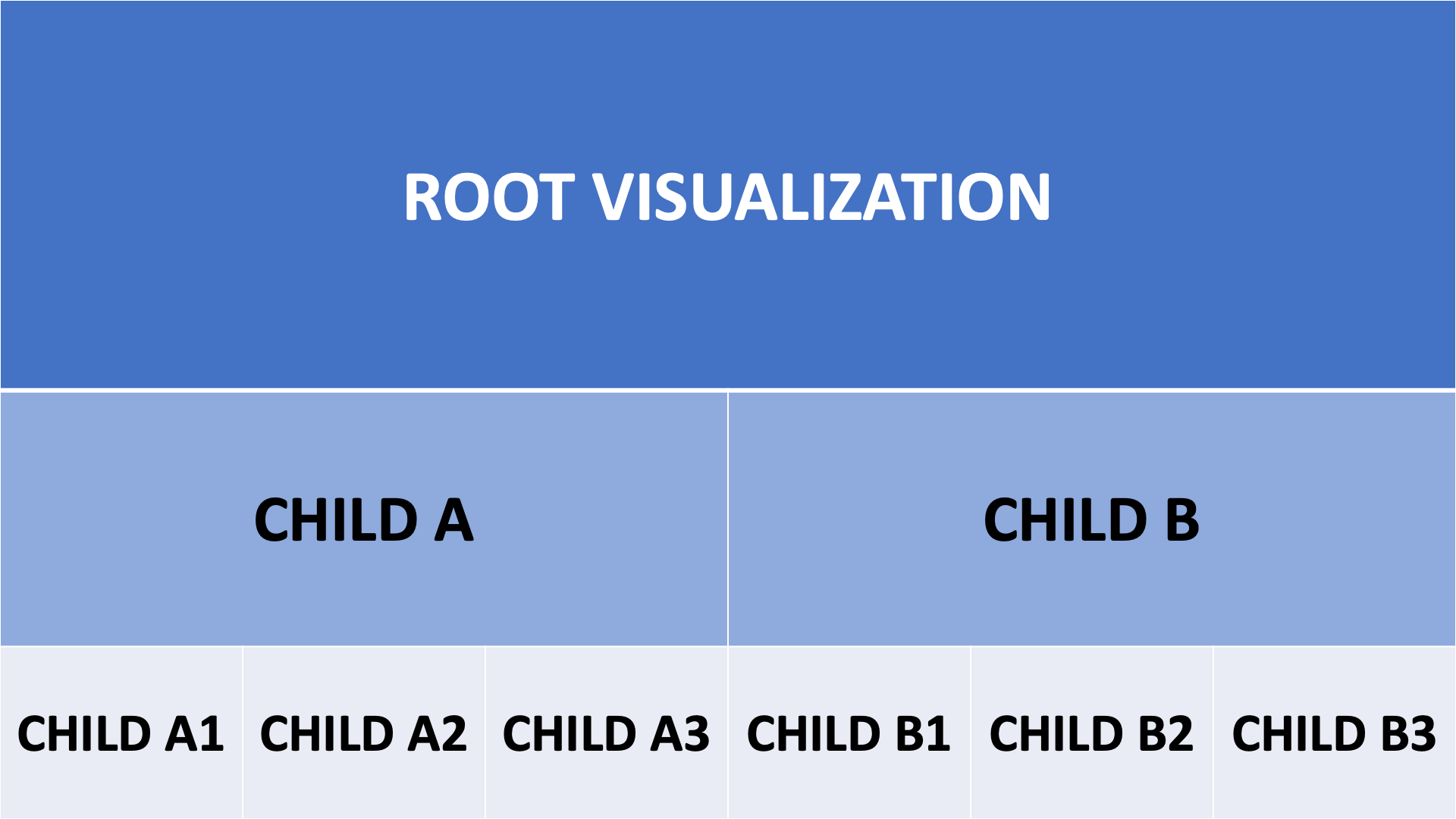}  
  \caption{Pyramidal Layout Style}
  \label{fig:pyramidal_style}
\end{subfigure} %
\begin{subfigure}[t]{.32\textwidth}
  \centering
  \includegraphics[width=\linewidth]{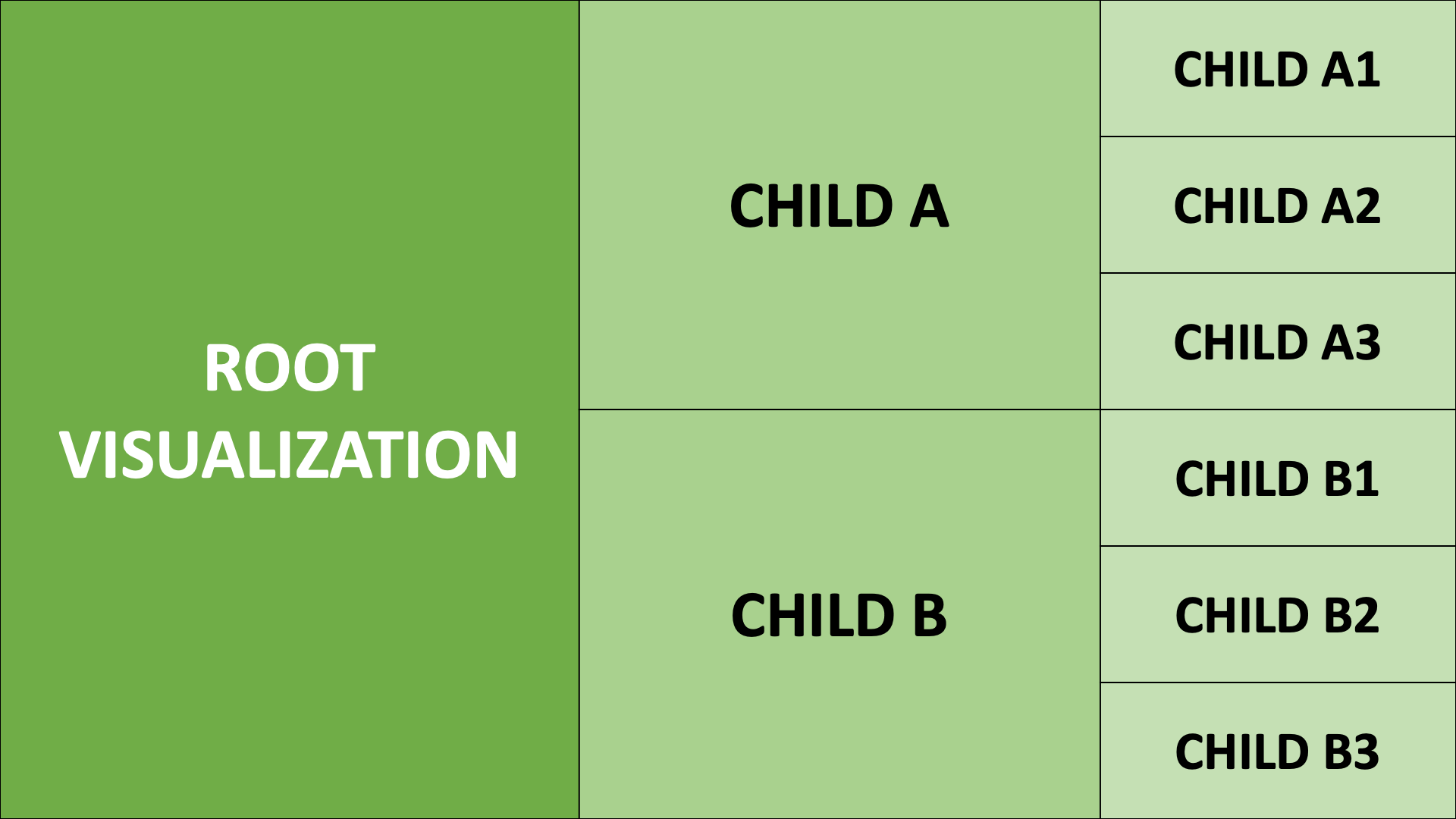}  
  \caption{Repeated Layout Style}
  \label{fig:repeated_style}
\end{subfigure}
\begin{subfigure}[t]{.32\textwidth}
  \centering
  \includegraphics[width=\linewidth]{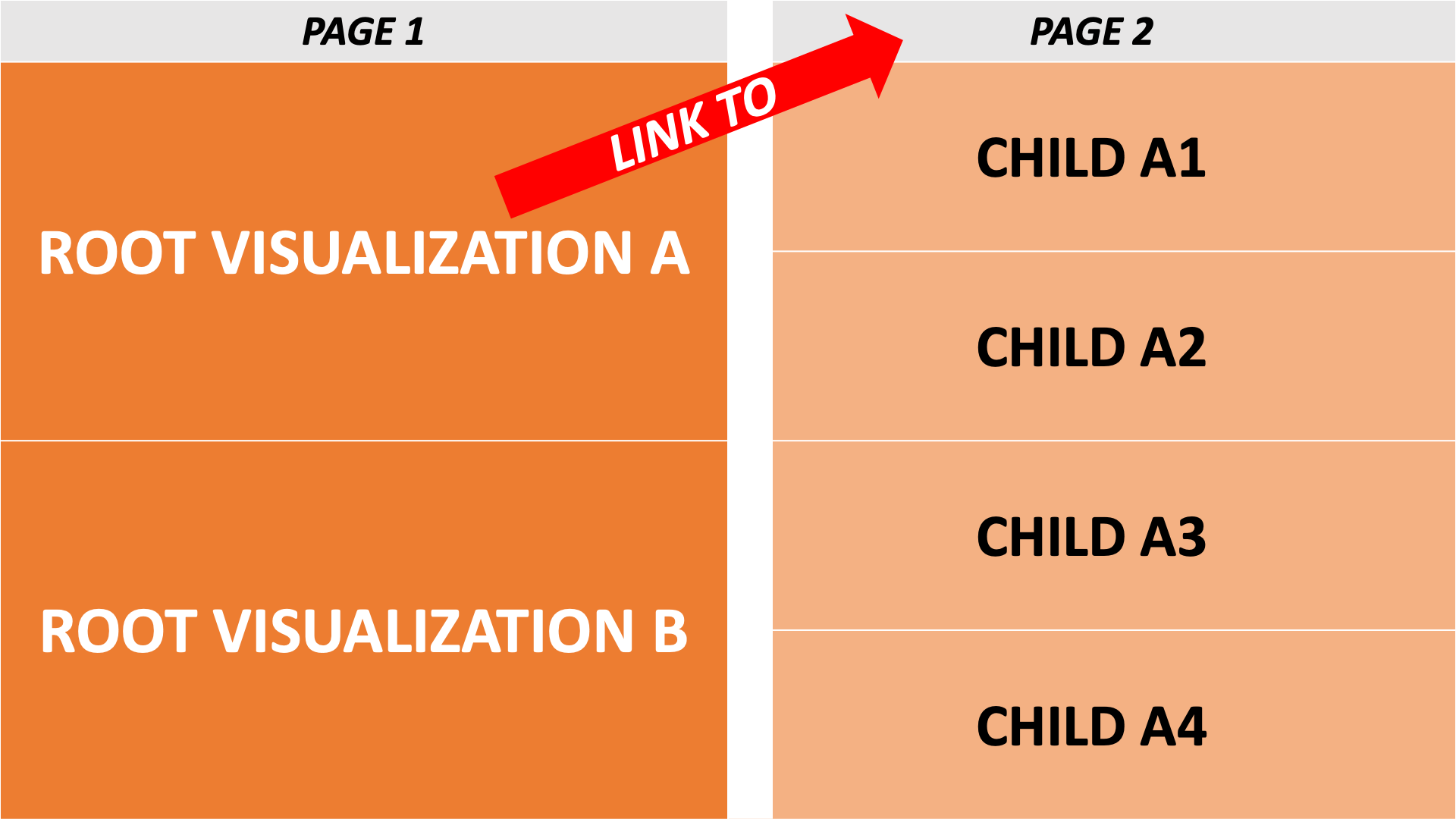}  
  \caption{Nested Layout Style}
  \label{fig:nested_style}
\end{subfigure}
\caption{Meta-Model Layouts}
\label{fig:metamodels}
\end{figure*}

In order to show these KPIs through Visualizations, both simple and composed visualizations can be used. A \textit{Simple Visualization} is defined as a set of KPIs that must be visualized together (see for example the definition of the CPU System visualization that shows the CPU System KPI in Listing \ref{lst:definition}, lines 14-16). Visualizations can be used to show the same indicator collected from different targets or different indicators collected from a same target in one visualization. In addition, visualizations may contain visualizations originating \textit{Composed Visualizations}. In particular, a Composed Visualization is defined by a set of Visualizations, either Simple or Composed, and by a \textit{Summary Visualization}, which is a Simple Visualization object that shows a summary overview of the composing elements at a glance (see for instance the CPU Composed Visualization defined in Listing~\ref{lst:definition}, lines 18-22). For instance, many visualizations presenting the health status of different resources can be summarized with a visualization that shows the resource with the worst health status.

\begin{lstlisting}[
caption={Declarative Dashboard Definition Example},
captionpos=b,
label={lst:definition},
language=yaml,
breaklines=true,
basicstyle=\footnotesize,
captionpos=b,                    
keepspaces=true,                 
numbers=right,                    
numbersep=-10pt,                  
showspaces=false,                
showstringspaces=false,
showtabs=false,                  
tabsize=2]
kpis:
	- name: CPU System
		metric: cpu_sys_pct
		target:
			id: websrv-01
			cluster: eu-west-01-dev
	[...]
	- name: CPU Total
		source_kpis:
			- CPU System
			- CPU User
		transformation_function: avg
visualizations:
	- name: CPU System
		kpis:
			- CPU System
	[...]
	- name: CPU 
		composing_visualizations:
			- CPU System
			- CPU User
		summary_visualization: CPU Overview
\end{lstlisting}

\subsection{Dashboard Transformation Based on Meta-model Layouts}
The result of the first step is the definition of a collection of KPIs and related Visualizations without a clear arrangement in a coherent layout. 
The second step exploits meta-models to wrap Visualizations into a new set of objects that define their arrangement in a dashboard layout. Note that the dashboard layout is still independent of the dashboard tool that is used to finally visualize the dashboard. Filling this gap defining how each element should be concretely rendered based  on the selected dashboard tool is the role of the third step.  

The meta-model has the sole responsibility of defining how to dispose the Visualization objects in a virtual dashboard that includes one or more pages each one including multiple spatially-arranged dashboard items. The meta-model also guides both the definition of the relative sizes of the visualizations and the navigation flow across pages.

More rigorously, a \textit{Virtual Dashboard} is a set of linked Dashboard Pages. A \textit{Dashboard Page} is a set of spatially-arranged Dashboard Items. A \textit{Dashboard Item} is a container of one or more related Visualizations.

So far, we defined two single-page meta-model layouts and a multi-page layout. The \textit{Pyramidal} and \textit{Repeated} meta-models generate a single-page Virtual Dashboard with Dashboard Items organized in a pyramid and in repeated groups, respectively. These two layouts differ from a visual perspective, but they are functionally equivalent. The \textit{Nested} meta-model generates a multi-page Virtual Dashboard with navigation links to browse across Visualizations. The selection of a specific meta-model can be done statically (e.g., using a configuration file), but we envision also the possibility to select the meta-model dynamically (e.g., using an API).

The Pyramidal meta-model maps each root Visualization, that is, each Visualization that is not included in another Visualization, at the top of a Dashboard Item. The root visualization spans the entire Dashboard Item. The composing visualizations, if any, are placed below the root visualization. A configurable number of visualizations is placed over multiple rows (if sliding is possible, animations can be used to place all the visualizations into a single row). If any of these visualizations is a Composed Visualization, the pattern is repeated. This is applied up to a maximum depth of three levels (again this limit can be configured). If an excessive number of levels is required to show all the visualizations, this style cannot be applied. Figure~\ref{fig:pyramidal_style} shows the logical arrangement of the visualizations when the Pyramidal layout style is applied.

The Repeated meta-model also maps all the visualizations into a single page. Differently from the Pyramidal style, the Repeated meta-model does not exploit the nested visualizations below the main visualization, but it shows visualizations from left to right. In particular, the composing visualizations are shown at the right of the composed visualization. This pattern is repeated for every composed visualization up to three levels of depth, otherwise the dashboard page is not generated.
Figure~\ref{fig:repeated_style} shows the logical arrangement of the visualizations when the Repeated layout style is applied.

The Nested meta-model maps visualizations across multiple pages. The first Dashboard Page includes all the root Visualizations. The visualizations included in a same root visualization are visualized in a dedicated Dashboard Page properly linked to the first page. All the visualizations have the same size in each page since they all belong to the same level of the hierarchy. This pattern is repeated for every composed visualization. Figure~\ref{fig:nested_style} shows the logical arrangement of the visualizations when the Nested layout style is applied.

\subsection{Concrete Dashboard Rendering}
The last step of the declarative dashboard generation produces the concrete input for a dashboard rendering tool, such as Grafana~\cite{grafanalabs2020grafana} and Kibana~\cite{elasticsearch2020kibana}, based on the virtual dashboard resulting from the application of the meta-model.

This step is tool-specific, in fact how a Visualization, a Dashboard Item, a Dashboard Page and the navigation links must be rendered depend on the target technology. Moreover, not every item of the Virtual Dashboard can necessarily be mapped into an item of the target dashboard tool. Finding the best compromise between the Virtual Dashboard and the rendered dashboard is a responsibility of this last step. 

We are currently working on the implementation of this mechanism for the most popular dashboard technologies.

\section{Related work}

Given the constantly increasing amount of data available from the monitoring and analysis of business processes, there is a growing need of useful data visualization and summarization technologies. Unfortunately, there is no single dashboard that can fit every purpose, so there is a need of creating \emph{tailored} dashboards for different contexts and users.

Vàzquez et al.~\cite{ingelmo2019survey} identify three macro types of tailored dashboards, namely \textit{Customized}, \textit{Personalized}, and \textit{Adaptive}. The main diversifying traits are \emph{when} the tailoring occurs and \emph{who} creates and performs the tailoring. 

Customized dashboards usually rely on explicit user requirements, therefore actively involving the users. The customization may happen through direct configuration of the dashboards or by exploiting configuration files or models that are manually produced.
As an example, many commercial solutions such as Kibana~\cite{elasticsearch2020kibana} and Grafana~\cite{grafanalabs2020grafana} offer features to customize dashboards, allowing users to create custom fitted dashboards using their graphical interfaces. However these solutions are limited since they always require the user intervention to create and reconfigure dashboards.
A more versatile example is the customizable dashboard system for microservices proposed by Mayer et al. in~\cite{mayer2017dashboard} that can be adapted for different stakeholders. This solution takes the heterogeneity and granularity of microservice-based systems into consideration, however, it cannot handle the dynamic situations that characterize the complex Systems of Systems considered in this paper.

Personalized dashboards usually infer their configurations from implicit data about the users or goals. It is noteworthy to clarify that in this context, the term \emph{goal} is used at different levels of abstractions: it may represent either a \emph{Monitoring Goal}, which is a high level description of an aspect that a user wants to monitor~\cite{shatnawi2018cloudhealth}, or a \emph{Business goal} as in the case of GQM~\cite{janes2013effective}, 
which is a domain-specific objective that must then be translated in one or more concrete monitoring goals.

Janes et al.~\cite{janes2013effective} propose a model-based personalized dashboard system with a particular focus on how a business goal can be expressed at different levels and how it evolves in more concrete monitoring goals and actual KPIs to be measured. They exploit this \emph{goal hierarchy} as a roadmap to define how the dashboard related to the business goal should look like, with a hierarchy of visualizations that matches the various levels of abstractions of the goal hierarchy. While their hierarchy of goals and visualizations is similar to the concept of Simple and Composed visualizations presented in this paper, the proposed dashboard system does not include features to create and automatically recreate and reconfigure the dashboard based on changes in the business goals. 

Kintz et al. \cite{kintz2017creating,kintz2012semantic} propose a similar business goal hierarchy, also describing a dashboard creation process based on VisML, a dedicated markup language that can be used to define the business goal hierarchy and the dashboard. 
VisML specifications are similar to the meta-model layouts presented in this paper. However the work by Kintz et al. \cite{kintz2017creating} relies on their own dashboard tool. To alleviate this limitation, our proposed dashboard generation approach completely decouples the meta-model layouts from the technology-dependent dashboards.

Continuously changing requirements and dynamic behaviors, such as the ones of large cloud-based Systems of Systems, generate the pressing need of having the ability to quickly and automatically reconfigure an available monitoring dashboard. The concept of \emph{adaptive dashboard} addresses these requirements by proposing a class of dashboards systems that can adapt themselves at runtime, based on environmental changes. Examples of adaptive dashboards are presented by Belo et al. in~\cite{belo2014restructuring} and Dabbebi et al. in~\cite{dabbebi2017towards}. The underlying concept of adaptivity presented in these articles pivots around the idea of tailoring not only which information is presented to different users, but also \emph{how} this information is presented to each user, based on preferences and past interaction with the systems. Also our meta-model layouts can support user-based customization by allowing the application of different meta-models to different users, based on their preferences. 


All the customized, personalized, and adaptive dashboard systems share some common downsides. As an example none of the mentioned solutions has been designed with Cloud-based Systems of Systems in mind and therefore they do not explicitly account for the frequent changes in the goals hierarchy typical of these kinds of systems. Moreover they rely on dashboard tools implemented ad-hoc for the specific solution. 

In this paper, we propose to overcome these issues by designing the dashboard generation process with a highly dynamic environment in mind, exploiting meta-models and a multi-step technology agnostic process, to define a general solution not strictly coupled with any dashboard tool. Our approach can fit Systems of Systems scenarios by enabling the individual operators to control their own dashboard definitions, exploiting the available monitoring data according to their rights and visibility of the system.

\section{Conclusions}

The complex and dynamic nature of Systems of Systems demands for highly-automated monitoring and data visualization solutions that can be quickly reconfigured to timely react to any change in the monitored KPIs. Existing solutions can address dashboard tailoring and reconfiguration, but they do not provide the level of automation and independence from dashboard tools required by modern software systems. 

This paper describes our early effort in the design of a largely automated dashboard generation and reconfiguration solution that is also agnostic from the specific dashboard rendering tool that is used. Our approach is originally driven by meta-model layouts that define how the final dashboard should look like. The presence of meta-models alleviates users from providing many details that are expensive to specify.

In the future, we plan to complete the design of the approach, providing a fully functional end-to-end solution for dashboard generation. Our vision is to achieve the capability to generate dashboards from the only knowledge of the collected KPIs.     

\vfill

\bibliographystyle{IEEEtran}
\bibliography{main}
\end{document}